\documentclass[]{raa}
\usepackage{graphicx,times}
\usepackage{natbib}

\begin{document}

   \title{Braking PSR J1734-3333 with a possible fall-back disk}

   \volnopage{Vol.0 (200x) No.0, 000--000}      
   \setcounter{page}{1}          

   \author{Xiong-Wei Liu\inst{1,2}, Ren-Xin Xu\inst{2}, Guo-Jun Qiao\inst{2}, Jin-Lin Han\inst{3}, and Hao Tong\inst{4}   }

   \institute{School of Physics, China West Normal University, Nanchong 637002, China; {\it xiongwliu@163.com}
   \and School of Physics and State Key Laboratory of Nuclear Physics and Technology, Peking University, Beijing 100871, China
   \and National Astronomical Observatories, Chinese Academy of Sciences, Beijing 100012, China
   \and Xinjiang Astronomical Observatory, Chinese Academy of Sciences, Urumqi 830011, China }

   \date{Received~~2013 month day; accepted~~2013~~month day}

\abstract{ The very small braking index of PSR J1734-3333,
$n=0.9\pm0.2$, challenges the current theories of braking mechanisms
in pulsars. We present a possible interpretation that this pulsar is
surrounded by a fall-back disk and braked by it. A modified braking
torque is proposed based on the competition between the magnetic
energy density of a pulsar and the kinetic energy density of a
fall-back disk. With this torque, a self-similar disk can fit all
the observed parameters of PSR J1734-3333 with natural initial
parameters. In this regime, the star will evolve to the region
having anomalous X-ray pulsars and soft gamma repeaters in the
$P-\dot{P}$ diagram in about 20000 years and stay there for a very
long time. The mass of the disk around PSR J1734-3333 in our model
is about $10M_{\oplus}$, similar to the observed mass of the disk
around AXP 4U 0142+61.  }

   \authorrunning{X. W. Liu et al. }            
   \titlerunning{Fall-back disk braking}  

   \maketitle

%
%
\section{Introduction}           
\label{sect:intro}

PSR J1734-3333 is a radio pulsar with a period $P$ = 1.17 s and a period derivative $\dot{P} = 2.28 \times 10^{-12}$. It is located between the normal radio pulsars and anomalous X-ray pulsars (AXPs) and soft gamma repeaters (SGRs) in the $P-\dot{P}$ diagram. PSR J1734-3333 has been observed regularly since 1997 by using the 64 m telescope at Parkes and the 76 m telescope at Jodrell Bank. It has not glitched during these years, insuring the accuracy of measuring $\nu$, $\dot{\nu}$ and $\ddot{\nu}$, where $\nu=1/P$ is the pulsar rotation frequency, $\dot{\nu}$ and $\ddot{\nu}$ are its first and second derivative. The braking index of a pulsar, $n\equiv \nu \ddot{\nu}/\dot{\nu}^2$, describes the dependence of the braking torque on rotation frequency. For PSR J1734-3333, $n=0.9\pm 0.2$, based on 13.5 years observational data \citep{Espinoza11}.

Because of the observational difficulties, only seven pulsars have reliable published braking indices \citep{Lyne93,Lyne96,Middleditch06,Livingstone07,Weltevrede11} before PSR J1734-3333. All of them have a value between 1 and 3 except PSR J0537-6910 which has glitched frequently \citep{Middleditch06}. The braking index value of PSR J1734-3333 is too small compared to that from spin-down models and the measured values of other pulsars. The pure magnetic braking with a constant dipolar magnetic field should have $n=3$. \cite{Xu01} considered the braking torques due to magnetodipole radiation and the unipolar generator, and predicted $1<n<3$. Other factors may also affect the braking index, such as the stellar wind, systematic variations of the moment of inertia or magnetic field, and non-dipolar braking \citep{Manchester85,Blandford88}. Therefore, the small braking index of PSR J1734-3333 is a challenge to the current braking mechanisms of pulsar.

From the definition of the braking index, one has $\dot{P}=k P^{2-n}$. When a pulsar evolves, its location on the $P-\dot{P}$ diagram will move according to $n$. The slope of the evolutionary path in the diagram is $2-n$. If PSR J1734-3333 keeps the present braking index value, it would dramatically evolve into the AXPs \& SGRs region in $\sim20,000$ yr. \cite{Espinoza11} suggested that PSR J1734-3333 may be a potential magnetar \citep{Thompson95,Thompson96},
whose magnetic field has been buried under the surface due to large accretion shortly after the supernova explosion, and is relaxing out of the surface at present. The increasing of magnetic field strength may also result in the small braking index value of 0.9.
In this paper, we propose another interpretation for the small braking index
that there may be a fall-back disk around PSR J1734-3333 which is braking the pulsar.


\section{Braking PSR J1734-3333 by a self-similar fall-back disk}
\subsection{The fall-back disk}

It is reasonable to assume that the mass could not be perfectly ejected during the supernova explosion and that a small amount of mass can fall back. A part of the fall-back material carry a sufficient angular momentum and can rotate around the young neutron star, forming a fall-back disk. The mass of a fall-back disk is unclear, but could be larger than $10M_{\oplus}$ and should be smaller than $0.1M_{\odot}$ \citep{Chevalier89,Lin91,Wang06}. PSR J1734-3333 is a young pulsar, with a characteristic age $\tau_c\sim8000$ yr. There may be a fall-back disk around it.

The fall-back disks are very difficult to be detected due to their extremely weak emission. About half of the AXPs \& SGRs have infrared/optical counterparts observed \citep{Wang02,Israel04,Israel05,Mereghetti11a}.
\cite{Wang06} observed the mid-infrared emission from a debris disk around AXP 4U 0142+61, the brightest known AXP. The disk mass is of the order of $\sim10M_{\oplus}$. \cite{Kaplan09} detected the counterpart to the AXP 1E 2259+586 at 4.5 $\mu$m, which can be explained by
the passive X-ray-heated dust disk model developed for 4U 0142+61.
On the theoretical point of view, the existence of disks around AXPs \& SGRs can help to understand their distribution on the $P-\dot{P}$ diagram, e.g. the origin of SGR 0418+5729 with a dipolar magnetic field less than $7.5 \times 10^{12}$ G \citep{van der Horst10,Rea10,Alpar11}. It is possible that PRS J1734-3333 will evolve to be a AXP or SGR, and now has a not-yet-detected disk around it.

When a fall-back disk is formed around a pulsar, the accretion rate will evolve over time. \cite{Cannizzo90} found that the accretion rate $\dot{M}$ declines self-similarly according to $\dot{M}\propto t^{-\alpha}$, where $\alpha>1$, when the disk is under the influence of viscous processes.
We parameterize the accretion rate in the same way as \cite{Chatterjee00}, i.e.
\begin{eqnarray}
& \dot{M}=\dot{M}_{\rm 0}, \hspace{1.2cm} & 0<t<T, \nonumber \\
& \dot{M}=\dot{M}_{\rm 0}(t/T)^{-\alpha}, & t\geq T,
\end{eqnarray}
where $\dot{M}_{\rm 0}$ is a constant accretion rate. \cite{Chatterjee00} supposed $T$ is the dynamical time in the inner parts of the disk, which is about 0.001 s. \cite{Menou01} pointed out that a more appropriate choice for $T$ is rather the viscous timescale, which is much larger than the dynamical time. We note that the choice for $T$ has no material effect for the spin evolution of a pulsar if one supposes a same value for $\dot{M}_{\rm 0}T^{\alpha}$. In this paper we take $T=1000$ s, to be order of the viscous timescale. With these assumptions, the total or initial disk mass is
\begin{equation}
M_{\rm d,0}=\int _0 ^{\infty} \dot{M} dt= \frac{\alpha}{\alpha-1}\dot{M}_{\rm 0}T,
\end{equation}
and the residual disk mass at time $t$, $t\geq T$, is
\begin{equation}
M_{\rm d}(t)=\int _t ^{\infty} \dot{M} dt= \frac{\dot{M}_{\rm 0}T^{\alpha}}{\alpha-1}t^{1-\alpha}.
\end{equation}
In this model, the super-Eddington accretion could occur at early
accretion phase. The initial parameters $\dot{M}_{\rm 0}$ and $T$
are coupling, and they are insensitive to the results at a late
stage. \cite{Cannizzo90} found $\alpha= 19/16$ and 1.25 for a disk
in which the opacity is dominated by electron scattering and for a
Kramers' opacity, respectively. We choose $\alpha= 7/6$, which is
very nearly the above values and could simplify the calculations, as
\cite{Chatterjee00} has shown. Our calculations show that small
changes of the $\alpha$ value do not effect the results
significantly.

\subsection{The braking torque}

The fall-back disk will interact with the pulsar magnetosphere, and influence the spin evolution of the pulsar, i.e.
\begin{equation}
I\dot{\Omega}=N,
\end{equation}
where $I$ and $\dot{\Omega}$ are the moment of inertia and angular velocity derivative of the pulsar,
and $N$ is the torque acting on the pulsar by the disk.
To study the spin evolution of a pulsar in detail, we introduce two important radii: one is the corotation radius $R_{\rm co}$ where the corotation velocity equals to the Kepler velocity, and the other is the magnetospheric radius $R_{\rm m}$ inside which the accretion disk cannot exist. The corotation radius
\begin{equation}
R_{\rm co}=(GM_*/4\pi^2)^{1/3}P^{2/3} \approx1.5\times10^8 M_1^{1/3}P^{2/3} \rm{cm},
\end{equation}
where $G$ is the gravitational constant, $M_*$ the mass of the star, and $M_1$ the mass of the star in units of solar mass. The magnetospheric radius
\begin{equation}
R_{\rm m}\approx 0.52r_{\rm A} = 8.7\times 10^7 \mu_{30}^{4/7}M_1^{-1/7}\dot{M}_{18}^{-2/7} \rm cm
\end{equation}
\citep{Ghosh79}, where $r_{\rm A}=\mu^{4/7}(2GM_*)^{-1/7}\dot{M}^{-2/7}$ is the Alfv\'en radius inside which the flow is dominated by the magnetic field, $\mu_{30}$ the magnetic moment of the star in units of $10^{30}$ G cm$^3$ (for a dipole magnetic field, $\mu=BR^3$), and $\dot{M}_{18}$ the accretion rate in units of $10^{18}$ g/s.

The sizes of $R_{\rm co}$ and $R_{\rm m}$ will change with the evolution of the pulsar and the disk, thereby the accretion process would undergo three phases successively: the accretion phase, the propeller phase and the tracking phase. At the early stage, when $R_{\rm m}<R_{\rm co}$, the accretion flow can fall onto the star surface, and the gravitational energy is released. In the accretion phase, $\dot{M}$ usually is larger than the Eddington limit ( $\dot{M}_{\rm E}\approx10^{18}$ g/s). In a spherically symmetric Eddington accretion case, the star would spin down very quickly braked by the strong coupling between the magnetic field and the ionized surroundings \citep{Liu12}. In the disk Eddington accretion case, the braking torque is still unclear. Fortunately, this stage is very short (in this paper, about several months to several years) and the spin period changes very little. Thus, we can simply ignore the effect of this stage and only suppose that the disk evolves self-similarly after the short initial accretion phase. The magnetospheric radius $R_{\rm m}$ increases faster than the corotation radius $R_{\rm co}$, and will exceed $R_{\rm co}$ after the accretion phase. When $R_{\rm m}>R_{\rm co}$, most of the accretion flow cannot reach the star surface, and the material would be ejected out by the propeller effect \citep{Illarionov75}. The propeller effect can brake the star to a relatively long spin period (typical several seconds), until $R_{\rm m}$ approximates to $R_{\rm co}$ and the propeller torque fades away. $R_{\rm m}$ could never equal to $R_{\rm co}$, but changes with $R_{\rm co}$. This is the tracking phase. In the whole evolution process, the propeller phase is the most important stage to brake the pulsar.

The propeller effect between a rotating magnetic neutron star and the accretion material was first proposed by \cite{Illarionov75}, to explain why the number of galactic X-ray sources is much less than the expected number of neutron stars and black holes in binary systems. They suggested that most of neutron star binary systems are in propeller phase, the accretion material is thrown away from the neutron star by the rotating magnetosphere, thus the stars could not emit enough X-ray luminosity to be detected. \cite{Manchester95} found evidence for a propeller-torque spin-down of PSR B1259-63 in a highly eccentric orbit binary system. Nevertheless, the value of propeller-torque is yet unclear.

The braking torque may be direct calculated by integrate the azimuthal magnetic stress in the coupling region. However, it has to make some very strong assumptions because there are too much uncertainties (see \cite{Ertan08,Caliskan12}). The torque may also be indirect estimated by calculate the angular momentum lost rate taken away by the ejected out material. For example, if the disk material rotates with Kepler angular frequency $\Omega_{\rm K}$ before it drops to $R_{\rm m}$, corotates with the star when it drops to $R_{\rm m}$ and is thrown away from the neutron star, it will take away angular momentum form the star, which equals to a braking torque
\begin{equation}
N=\dot{J}=2\dot{M}R_{\rm m}^2\Omega_{\rm K}(R_{\rm m})[1-\frac{\Omega}{\Omega_{\rm K}(R_{\rm m})}],
\end{equation}
\citep{Menou99,Chatterjee00}, where $\dot{J}$ is angular momentum exchange rate and $\Omega$ the angular frequency of the star.
In this formula, \rm the torque is zero when $\Omega$ equals to $\Omega_{\rm K}(R_{\rm m})$, and would increase with the difference between $\Omega$ and $\Omega_{\rm K}(R_{\rm m})$. This is reasonable and feasible. However, according to this formula, the kinetic energy density of the corotating material at $R_{\rm m}$ would become much larger than the magnetic energy density there when $\Omega >> \Omega_{\rm K}(R_{\rm m})$, which seems to be more or less unnatural. Considering both the reasonable and the unnatural factors, we modify the propeller torque with an additional parameter $\chi$, i.e.
\begin{equation}
N=2\dot{M}R_{\rm m}^2\Omega_{\rm K}(R_{\rm m}) \{1-[\frac{\Omega}{\Omega_{\rm K}(R_{\rm m})}]^{\chi}\},
\end{equation}
where the power exponent $\chi$ is smaller than one, i.e. $0<\chi<1$. This torque is more moderate than those with $\chi\geq1$ \citep{Chatterjee00,Ertan08}. We expect the torque with $0<\chi<1$ fit the observation better than that with $\chi\geq1$.

\subsection{The results}

\begin{figure}
  \centerline{\includegraphics[width=0.8\linewidth]{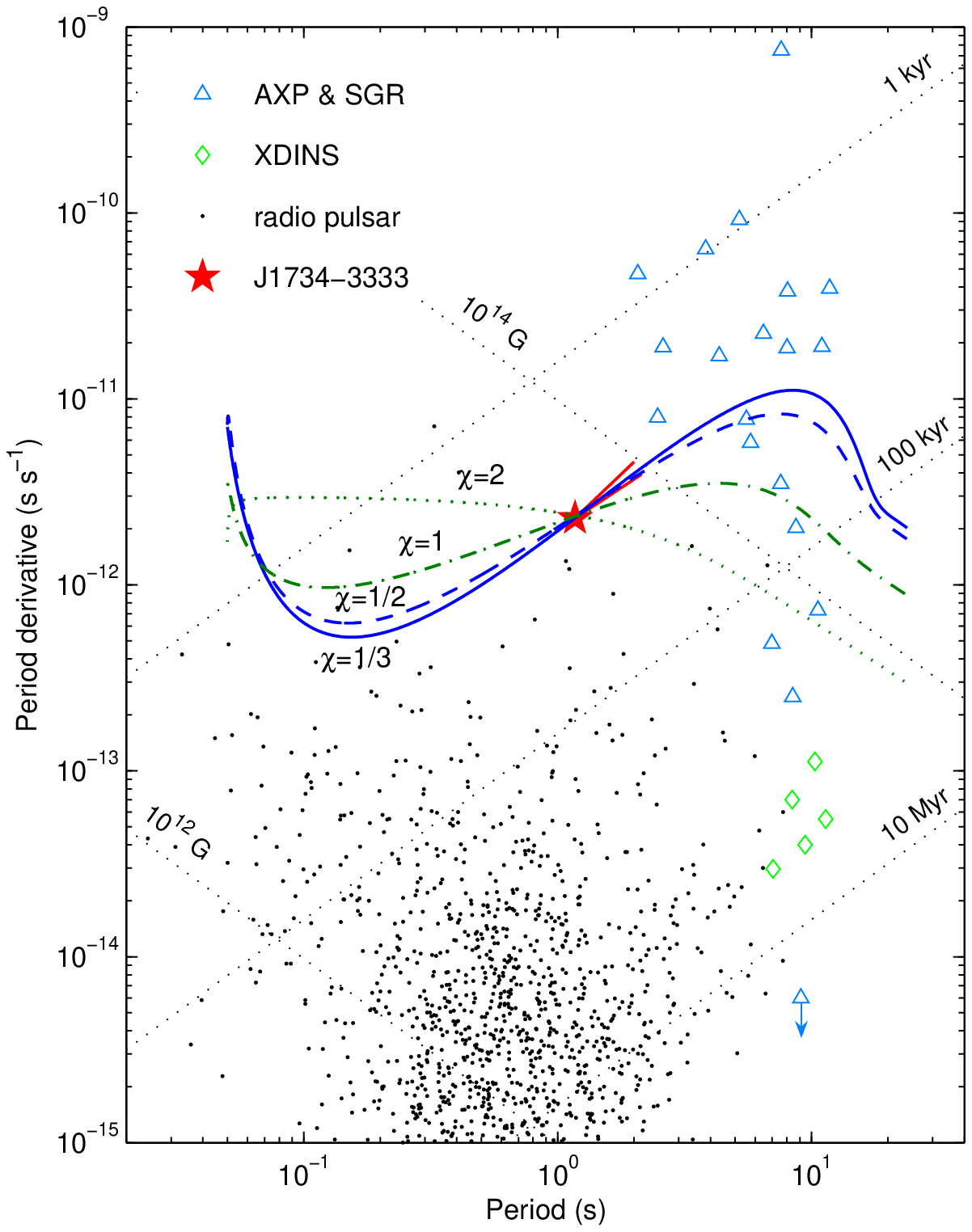} }
  \caption{Evolution paths of a neutron star surrounded by a fall-back disk with different propeller torques, given initial parameters $P_0=50$ ms, $T=1000$ s, and $\alpha=7/6$. To cross the location of PSR J1734-3333 in the diagram, different $\chi$ needs different disk mass and magnetic field strength. The fitting parameters and main results are given in Table 1. The two short lines near the star indicate the observational upper and lower limits of the braking index of PSR J1734-3333. The information of radio pulsars was taken from the ATNF pulsar catalogue (\cite{Manchester05}, http://www.atnf.csiro.au/research/pulsar/psrcat/), data of AXPs and SGRs from the McGill SGR/AXP Online Catalog (http://www.physics.mcgill.ca/$\sim$pulsar/magnetar/main.html), and X-ray Dim Isolated Neutron Stars from \cite{Mereghetti11b}. \label{fig1}}

\end{figure}

The period evolution of a pulsar can be calculated using equations (4) and (6), given the accretion history from equation (1) and the torque formula (8). The magnetic dipole radiation torque is insignificant compared with the propeller torque, especially when it spins slowly. We thus ignore the magnetic dipole radiation in the following calculations.

Figure 1 shows the evolution paths in the $P-\dot{P}$ diagram of a neutron star surrounded by a fall-back disk, with initial parameters $P_0=50$ ms, $M_1=1.4$, $T=1000$ s, and $\alpha=7/6$. We tried $\chi=1/3$ and $\chi=1/2$, whose paths can go across the location of PSR J1734-3333, and the braking indexes coincide with the observed value $n=0.9\pm 0.2$ \citep{Espinoza11}. For comparison, the case of $\chi=1$ \citep{Chatterjee00} and the case of $\chi=2$ \citep{Ertan08} are also calculated and shown in Figure 1, in which the braking indexes can not be equal to the observed value of PSR J1734-3333. As Figure 1 shows, the propeller phase can be divided into three stages according to the slope of the evolution path in the $P-\dot{P}$ diagram. In the first stage, the path goes down quickly until the slope increases from very negative to zero. In the second stage the path goes up, where the slope increases from zero to a value larger than 1, i.e. $n<1$, and decreases to zero at the end. In the last stage the path goes down again, and the slope keeps decreasing, until the neutron star evolves to the tracking phase \citep{Chatterjee00}. The fitting parameters and main results are given in Table 1. All the parameters and results vary with $\chi$ regularly.

It is difficult to judge whether $\chi=1/2$ or $\chi=1/3$ is the most likely solution, though it seems that $\chi=1/3$ fits the braking index better, because the real disk and the braking torque are unlikely as simple as the models. Nevertheless, the torques with $0<\chi<1$ fit the observation quite better than those with $\chi\geq1$, agree with our expectation when we propose $\chi$ to modify the torque.

\begin{table}
\bc
\caption{Fitting parameters and main results.\label{tbl-2}}
\begin{tabular}{crrrrrrrrrrr}
\hline\noalign{\smallskip}
$\chi$ & $\mu_{30}$ & $M_{\rm d,0}(M_{\oplus})$ & $P$ (s) & $\dot{P}$ &  $n$ & $t(\rm{kyr})$ & $M_{\rm d}(t)(M_{\oplus})$\\
\hline\noalign{\smallskip}
1/3 &5.8 &525 &1.17& $2.28\times10^{-12}$ &0.97 &34.2 &14 \\
1/2 &4 &292 &1.17& $2.28\times10^{-12}$ &1.10 &30.2 &8 \\
1 &1.7 &117 &1.17& $2.28\times10^{-12}$ &1.52 &22.9 &3.3 \\
2 &0.8 &93 &1.17& $2.28\times10^{-12}$ &2.25  &13.4 &2.9 \\
\noalign{\smallskip}\hline
\end{tabular}
\ec
\tablecomments{0.9\textwidth}{ $\mu_{30}$ and $M_{\rm d,0}$ are the only two free fitting parameters. $P$, $\dot{P}$, $n$, $t$ and $M_{\rm d}(t)$ are the fitted results when the pulsar passes the location of PSR J1734-3333. }
\end{table}

\begin{figure}
\centerline{\includegraphics[width=0.8\linewidth]{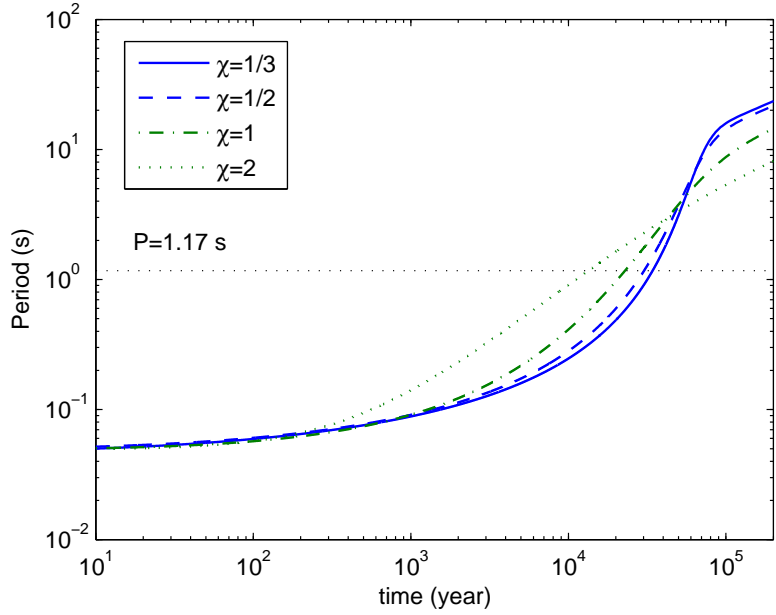} }
\caption{Period evolution of a neutron star with $P_{\rm 0}=50$ ms surrounded by a disk with different parameters given in Table 1. \label{fig2}}
\end{figure}

\begin{figure}
\centerline{\includegraphics[width=0.8\linewidth]{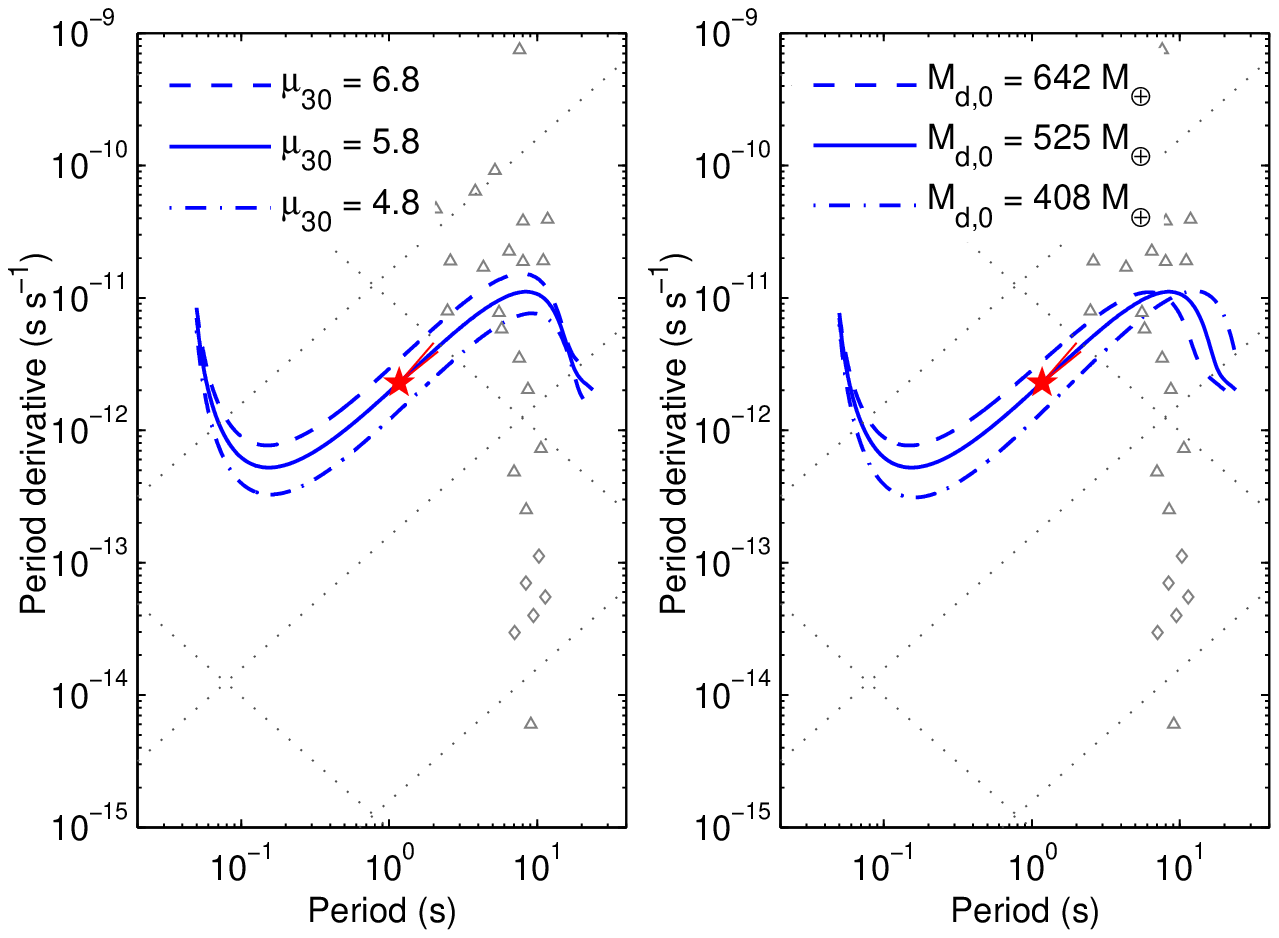} }
\caption{Evolution paths of a neutron star with different magnetic field strength and disk mass, which braked by a disk torque with $\chi=1/3$. The same parameters are $P_0=50$ ms, $M_1=1.4$, $T=1000$ s and $\alpha=7/6$. Left panel: $M_{d,0}=525 M_{\oplus}$; Right panel: $\mu_{30}=5.8$. \label{fig3}}
\end{figure}

Figure 2 shows the period evolution with the different parameters given in Table 1. The period of a neutron star surrounded by a fall-back disk could evolve to several seconds in about tens of thousands of years and could stay in the period range for a very long time.
Figure 3 shows the evolution paths of a pulsar with different magnetic field strengths and initial disk masses, where $\chi=1/3$. The paths are similar and change with the parameters steadily. In the second stage, the slopes of the paths vary little and are all similar to the observation of PSR J1734-3333, which means that the braking indexes of pulsars in propeller phase with self-similar fall-back disks are all about 1.

\section{Conclusion and Discussion}

In this paper we proposed a modified formula for the propeller torque considering the energy densities of the magnetosphere and the accretion material, motivated by previous works \citep{Menou99,Chatterjee00,Alpar01,Ertan08}. With such a torque, a self-similar fall-back disk can make the braking index of a pulsar very small, which fits the observed $P$, $\dot P$ and $n$ of PSR J1734-3333 very well. The results are stable and insensitive to the initial parameters.

Shortly after we presented our work on arXiv (arXiv:1211.4185), \cite{Caliskan12} presented another fall-back disk solution for the small braking index of PSR J1734-3333. However, the torque used in their paper is doubtable. \cite{Ertan08} obtained the torque based on at lest two assumptions, the disk material extends to the corotation radius $R_{\rm co}$ and the azimuthal pitch of the magnetic field in the region between $R_{\rm co}$ and $r_{\rm A}$ is constant, which are somewhat unreasonable. As a result, \cite{Caliskan12} needs a small disk whose mass is much smaller than the mass of the disk around AXP 4U 0142+61 \citep{Wang06} to fit the observational parameters of PSR J1734-3333, and the fitting is very sensitive to the initial conditions. (see figure 4 of \cite{Caliskan12} and figure 3 of this paper for comparison)

As shown in Table 1, the initial disk masses are smaller than $0.1M_{\odot}$ which agrees with \cite{Chevalier89} and \cite{Lin91}. When PSR J1734-3333 evolves into the AXP \& SGR region in about 20,000 years later, the disk mass would be of the order of $10M_{\oplus}$, similar to the observed disk mass around AXP 4U 0142+61 \citep{Wang06}. The propeller torque of a fall-back disk would modify the period derivative, which makes the apparent dipole magnetic field strength much stronger than the real field strength. Our model implies that the magnetic field strength of PSR J1734-3333 is only about $10^{12}$ G, much smaller than the value expected from the pure magnetic braking.

As figures 1 and 3 show, the evolution pathes sharply go down in the last stage, when the pulsar evolves into the region of AXPs \& SGRs in the $P-\dot{P}$ diagram. At that time, the inner disk is very close to the corotation radius, which makes it be possible that a large part of the inner disk material pass through the corotation radius and fall onto the neutron star surface. Thus the star may act as an AXP then, just like \cite{Chatterjee00} and \cite{Alpar01} have suggested. If this pulsar really becomes an accretion powered AXP in the future, the braking torque will be more complex than the torque used in our model, thus it will be more difficult to predict the evolution path of a pulsar in the region of AXPs \& SGRs. We leave this issue to the future work since this paper devotes mainly to explain the small braking index.

\begin{acknowledgements}

We would like to thank the pulsar group of PKU for useful discussions.
This work is supported by the National Basic Research Program of China (Grant Nos. 2009CB824800, 2012CB821800), the National Natural Science Foundation of China (Grant Nos. 11225314, 10935001, 11103021, 11373011), and XTP project XDA04060604.
\end{acknowledgements}


\begin{thebibliography}{}

\bibitem[Alpar(2001)]{Alpar01} Alpar, M.~A.\ 2001, \apj, 554, 1245

\bibitem[Alpar et al.(2011)]{Alpar11} Alpar, M.~A., Ertan, {\"U}., \& {\c C}al{\i}{\c s}kan, {\c S}.\ 2011, \apjl, 732, L4

\bibitem[Blandford \& Romani(1988)]{Blandford88} Blandford, R.~D., \& Romani, R.~W.\ 1988, \mnras, 234, 57P

\bibitem[{\c C}ali{\c s}kan et al.(2013)]{Caliskan12} {\c C}ali{\c s}kan, {\c S}., Ertan, {\"U}., Alpar, M.~A., Tr{\"u}mper, J.~E.,
\& Kylafis, N.~D.\ 2013, \mnras, 842, arXiv:1211.4689

\bibitem[Cannizzo et al.(1990)]{Cannizzo90} Cannizzo, J.~K., Lee, H.~M., \& Goodman, J.\ 1990, \apj, 351, 38

\bibitem[Chatterjee et al.(2000)]{Chatterjee00} Chatterjee, P., Hernquist, L., \& Narayan, R.\ 2000, \apj, 534, 373

\bibitem[Chevalier(1989)]{Chevalier89} Chevalier, R.~A.\ 1989, \apj, 346, 847

\bibitem[Ertan \& Erkut(2008)]{Ertan08} Ertan, {\"U}., \& Erkut, M.~H.\ 2008, \apj, 673, 1062

\bibitem[Espinoza et al.(2011)]{Espinoza11} Espinoza, C.~M., Lyne, A.~G., Kramer, M., Manchester, R.~N., \& Kaspi, V.~M.\ 2011, \apjl, 741, L13

\bibitem[Ghosh \& Lamb(1979)]{Ghosh79} Ghosh, P., \& Lamb, F.~K.\ 1979, \apj, 234, 296

\bibitem[Illarionov \& Sunyaev(1975)]{Illarionov75}Illarionov, A. F., \& Sunyaev, R. A. 1975, A\&A, 39, 185

\bibitem[Israel et al.(2005)]{Israel05} Israel, G.~L, Covino, S., Mignani, R., et al.\ 2005, \aap, 438, L1

\bibitem[Israel et al.(2004)]{Israel04} Israel, G.~L., Rea, N., Mangano, V., et al.\ 2004, \apjl, 603, L97

\bibitem[Kaplan et al.(2009)]{Kaplan09} Kaplan, D.~L., Chakrabarty, D., Wang, Z., \& Wachter, S.\ 2009, \apj, 700, 149

\bibitem[Lin et al.(1991)]{Lin91} Lin, D.~N.~C., Woosley, S.~E., \& Bodenheimer, P.~H.\ 1991, \nat, 353, 827

\bibitem[Liu et al.(2012)]{Liu12} Liu, X.~W., Xu, R.~X., Qiao, G.~J., et al.\ 2012, (submitted to ApJ) arXiv:1207.4687

\bibitem[Livingstone et al.(2007)]{Livingstone07} Livingstone, M.~A., Kaspi, V.~M., Gavriil, F.~P., et al.\ 2007, \apss, 308, 317

\bibitem[Lyne et al.(1993)]{Lyne93} Lyne, A.~G., Pritchard, R.~S., \& Graham-Smith, F.\ 1993, \mnras, 265, 1003

\bibitem[Lyne et al.(1996)]{Lyne96} Lyne, A.~G., Pritchard, R.~S., Graham-Smith, F., \& Camilo, F.\ 1996, \nat, 381, 497

\bibitem[Manchester et al.(2005)]{Manchester05} Manchester, R.~N., Hobbs, G.~B., Teoh, A., \& Hobbs, M.\ 2005, \aj, 129, 1993

\bibitem[Manchester et al.(1995)]{Manchester95} Manchester, R.~N., Johnston, S., Lyne, A.~G., et al.\ 1995, \apjl, 445, L137

\bibitem[Manchester et al.(1985)]{Manchester85} Manchester, R.~N., Newton, L.~M., \& Durdin, J.~M.\ 1985, \nat, 313, 374

\bibitem[Menou et al.(1999)]{Menou99} Menou, K., Esin, A.~A., Narayan, R., et al.\ 1999, \apj, 520, 276

\bibitem[Menou et al.(2001)]{Menou01} Menou, K., Perna, R., \& Hernquist, L.\ 2001, \apj, 559, 1032

\bibitem[Mereghetti(2011a)]{Mereghetti11a} Mereghetti, S.\ 2011a, Advances in Space Research, 47, 1317

\bibitem[Mereghetti(2011b)]{Mereghetti11b} Mereghetti, S.\ 2011b, High-Energy Emission from Pulsars and their Systems, 345

\bibitem[Middleditch et al.(2006)]{Middleditch06} Middleditch, J., Marshall, F.~E., Wang, Q.~D., Gotthelf, E.~V., \& Zhang, W.\ 2006, \apj, 652, 1531

\bibitem[Rea et al.(2010)]{Rea10} Rea, N., Esposito, P., Turolla, R., et al.\ 2010, Science, 330, 944

\bibitem[Thompson \& Duncan(1995)]{Thompson95} Thompson, C., \& Duncan, R.~C.\ 1995, \mnras, 275, 255

\bibitem[Thompson \& Duncan(1996)]{Thompson96} Thompson, C., \& Duncan, R.~C.\ 1996, \apj, 473, 322

\bibitem[van der Horst et al.(2010)]{van der Horst10} van der Horst, A.~J., Connaughton, V., Kouveliotou, C., et al.\ 2010, \apjl, 711, L1

\bibitem[Wang et al.(2006)]{Wang06} Wang, Z., Chakrabarty, D., \& Kaplan, D.~L.\ 2006, \nat, 440, 772

\bibitem[Wang \& Chakrabarty(2002)]{Wang02} Wang, Z., \& Chakrabarty, D.\ 2002, \apjl, 579, L33

\bibitem[Weltevrede et al.(2011)]{Weltevrede11} Weltevrede, P., Johnston, S., \& Espinoza, C.~M.\ 2011, \mnras, 411, 1917

\bibitem[Xu \& Qiao(2001)]{Xu01} Xu, R.~X., \& Qiao, G.~J.\ 2001, \apjl, 561, L85


\end{thebibliography}
\end{document}